\documentstyle[12pt,aasms4]{article} 
\newcommand{\asca}{{\it ASCA} }

\newcommand{\chisq}{$\chi ^{2}$ }
\newcommand{\etal}{{\it et al.} }

\accepted{16 January 1998}
\begin{document}

\title{PARAMETER ESTIMATION IN X-RAY ASTRONOMY REVISITED} 

\vspace{2cm}
 
\author{Tahir Yaqoob\altaffilmark{1}}
\affil{ Laboratory for High Energy Astrophysics, 
NASA / Goddard Space Flight Center,} 
\centerline{Code 662, Greenbelt, MD 20771.}
\altaffiltext{1}{ Also with the Universities Space Research Association.}

\vspace{4cm}
\begin{center}
NOTE: This version supercedes all previous versions you may have seen.
Substantially incorrect versions of this paper were sent out due
to a departmental clerical error.
\end{center}
\vspace{1cm}

\centerline{\it To appear in the Astrophysical Journal, June 20, 1998, vol. 499}

\begin{abstract} 
The method of obtaining confidence intervals
on a subset of the total number of parameters ($p$) of a model
used for fitting X-ray spectra is to perturb the best-fitting model
until, for each parameter, a range is found for which the
change in the fit statistic is equal to some critical value.
This critical value corresponds to the
desired confidence level and is obtained from the $\chi^{2}$ distribution
for $q$ degrees of freedom, where $q$ is the number of interesting
parameters. 
With the advent of better energy-resolution detectors, such
as those onboard {\it ASCA} 
({\it Advanced Satellite for Cosmology and Astrophysics})
it has become more common to fit complex
models with narrow features, comparable to the instrumental energy resolution.
To investigate whether this leads to significant non-Gaussian deviations
between data and model, we  
use simulations based on \asca 
data and we show that
the method is still valid in such cases. 
We also investigate the weak-source limit as well as the case of  
obtaining upper limits on equivalents widths of weak
emission lines and find that for all practical purposes the method
gives the correct confidence ranges. However, upper limits on emission-line
equivalent widths may be over-estimated in the extreme Poisson limit.

\end{abstract}
 
\keywords{ methods: data analysis -- methods: statistical -- 
X-rays: general }

\section{ INTRODUCTION} 
\label{intro}

The procedure for generating confidence intervals for the parameters of
models used to fit X-ray spectra with the \chisq  minimization
technique is well established (e.g. Lampton, Margon and Bower 1976;
Avni 1976).
One uses the fact that if $\chi ^{2} _{\rm true}$ is the value of
the statistic from a given experiment, calculated for the `true model' (known
only to nature) and $\chi_{\rm min}^{2}$ is the value of the statistic
obtained from the best-fitting model, then from many repetitions of the same
experiment $\Delta \chi^{2} = \chi^{2}_{\rm true} - \chi^{2}
_{\rm min}$ has a probably distribution like $\chi ^{2}$ with $p$
degrees of freedom ($\equiv \chi^{2}_{p}$)
where $p$ is the total number of free parameters.
Confidence intervals generated for the $p$
parameters from the $\chi^{2}_{p}$ distribution are then {\it joint}
confidence intervals for all $p$ parameters. 
If we are interested only in a subset $q$ of the $p$
parameters then the confidence intervals are generated from
the $\chi^{2}_{q}$ distribution. The crucial difference between
$p$-parameter and, say, one-parameter errors is as follows.
In the former case the $P\%$ confidence intervals
from $\chi^{2}_{p}$ will {\it simultaneously} enclose the true values of all
parameters in $P\%$ of all experiments. In the latter case,
the $P\%$ intervals from $\chi^{2}_{1}$ 
will enclose the true values
of any of the parameters in $P\%$ of all experiments, 
{\it but it will be a different $P\%$ subset of experiments for each
parameter}.
In a particular case, the number of `interesting' parameters (i.e. $q$)
is determined by the scientific problem being posed.

In principle one generates the $\Delta \chi ^{2}$ space by
stepping through a $q$-dimensional grid of parameter values.
The use of $\chi ^{2}$ in model-fitting requires that there are a sufficient
number of 
photons per energy bin 
for the statistical variations to be
Gaussian. 
However, with the advent of better
energy resolution X-ray detectors it is increasingly becoming the case that
the Gaussian approximation cannot be made unless the spectrum is binned,
sacrificing valuable information. In such cases the statistical variation in
counts per bin is Poisson and one must use the maximum likelihood ratio
(hereafter, `$C$ statistic')
to optimize the model parameters
(see Cash 1979). Cash (1979) demonstrated that
the $C$ statistic can be used to generate confidence intervals in an analogous
manner to $\chi ^{2}$ since $\Delta C$ has a probability distribution like
$\chi ^{2}$ except for terms of order $\alpha/ \sqrt{n}$ where $\alpha$
depends on the model and $n$ is the number of photons carrying information
about the parameter in question. 
Hereafter, we shall use $C$ for the sake of generality.

Lampton \etal (1976) warned that (at that time),
there was no general proof that the technique for projection of 
the subset of $q$ parameters did not
depend explicitly on model linearity.
Then, Avni (1976), using some non-linear models in  simulations of
{\it Uhuru} data, showed that the $\chi^{2}_{q}$ region worked
in these particular cases, but there was still no general proof.
Such a proof was presented by Cash (1976), 
showing that the $\chi^{2}_{q}$ region worked for any data set, 
{\it provided that the deviations are
Gaussian}. 
We must remember that
X-ray detectors now have much better energy resolution and sensitivity
than they did then and accordingly the models have become much more
complex. Both Lampton \etal (1976) and Avni (1976) used 
{\it Uhuru} spectral responses
with seven energy bins between 1 and 7 keV, and simple power-law
or plasma models. 
It is not clear whether, for example, a model including a narrow
emission line whose intrinsic width is comparable to the energy 
resolution of the detector, would introduce non-Gaussian
deviations between model and data. 
The purpose of this paper is to 
check whether the method for parameter estimation currently in use gives the
correct confidence intervals even 
with
the new generation of improved energy-resolution instrumentation.
We are particularly interested in 
models with 
features such as emission
lines and absorption edges (which occur
frequently in a wide range of X-ray sources), whose widths in energy space are
comparable to the instrumental energy resolution. We also wish to
check the Poisson regime, when the source count rate is low, and when
upper limits must be obtained on the equivalent
width of weak or non-detected emission-line features. 
 
The structure of the paper is as follows:
\S \ref{simulations} describes basic simulations
and models used in the investigation; \S \ref{results} describes the
basic results;
\S \ref{lineew} demonstrates the equivalence of emission-line
equivalent width and intensity confidence regions; \S \ref{weaksource}
describes results for the extreme Poisson limit and \S 
\ref{weakline} describes results pertaining to measuring upper limits
on weak emission-line features. 
Our conclusions are stated in \S \ref{conclusions}.

\section{SIMULATIONS}
\label{simulations}

We investigated the behavior of $\Delta C$
for various models
by means of simulations of spectra
from {\it ASCA} (Tanaka, Inoue and Holt 1994), using the spectral
fitting code XSPEC (Shafer \etal 1989). 
The spectral response function for one of the `Solid-state Imaging
Spectrometers' aboard \asca (SIS0) was used in the simulations. 
The simulated spectra consisted  of
330 pulse height bins of width $\sim 0.03-0.3$ keV covering the 
energy range $0.5-10$ keV. At 6 keV the energy
resolution is $\sim 2\%$, or $\sim 130$ eV.
No attempt was made to simulate the internal or sky background.
In practice, if one is using the $C$ statistic the background
cannot be subtracted since $C$ strictly requires Poisson statistics. The
background must be modelled and the resulting model included together
with the source model in the spectral fitting process, fixing the
background model parameters, which therefore
do not make any contribution to $\Delta C$.

We used six different models, described in Table 1, where
the exact parameter values are specified.
In each case the basic continuum model
is a power law plus absorption, typical of the X-ray spectra of active
galactic nuclei (AGNs). The 
2--10 keV flux corrected for absorption was $\sim 5 \times 10^{-11} \rm
\ ergs \ cm^{-2} \ s^{-1}$ and the 
exposure time was 40 Ks in each case (unless specified
otherwise). This exposure time is typical
for \asca observations.
For each model the following process was  
repeated 1000 times. A simulated spectrum was  
created and the value of the statistic recorded. This is just 
$C_{\rm true}$, which characterizes the Poisson deviations of the
data away from the true (i.e input) model. A spectral fit was  then performed
on the simulated data using the same model, with all the parameters
free. The best-fitting parameters were   recorded, as well as the 
corresponding value of the statistic, $C_{\rm min}$. 
Next, each parameter, $i$, was fixed at its true (input) value
and then $C$ was minimized over the remaining parameters, giving
a value, $C_{i}$, for each parameter. We then examined 
the probability distributions of $\Delta C_{\rm true} \equiv C_{\rm true}
- C_{\rm min}$ and $\Delta C_{i} \equiv C_{i} - C_{\rm min}$, as well as
the behavior of $\Delta C_{\rm true}$ and
$\Delta C_{i}$ as a function of the best-fitting parameter values. 
We did not always use the results of all 1000 simulations
since the few cases in which any of the $\Delta C$ were negative
were rejected since this indicated that a proper minimization had
not been achieved.

\section{ BASIC RESULTS}
\label{results}

Figures 1a--1f show the cumulative probability distributions of 
$\Delta C_{\rm true}$ (stars) and the $\Delta C_{i}$
(symbols used for the different parameters are 
explained in the Figures). The solid lines show the
theoretical $\chi^{2}$ distributions for 1 to $p$ degrees of freedom
where $p$ is the total number of free parameters.
Model 1 is the `control' case, consisting only of a simple
power law plus absorption with no localized, or narrow, features. 
In this case,
Figure 1a shows that $\Delta C_{\rm true}$ is
distributed like 
$\chi_{3} ^{2}$ as expected, since  $p=3$. 
Thus, for a given
$\Delta C_{\rm true}$ and corresponding percentage probability, $P$,
the $P\%$ {\it joint} confidence intervals of all $p$ parameters will
be given by the range of parameters associated with all the
simulations that have $\Delta C \le \Delta C_{\rm true}$. 
This is because $\Delta C_{\rm true}$ is computed with $p$
adjustable parameters and $P\%$ of {\it all} experiments
have $\Delta C \le \Delta C_{\rm true}$.
Figure 1a also shows that all the $\Delta C_{i}$ are correctly
distributed as $\chi_{1} ^{2}$.  The ranges of 
parameters associated with these distributions
correspond  to confidence intervals for one interesting parameter.
For a given $\Delta C_{i}$ and the corresponding percentage probability, $P$,
we can say that in $P\%$ of all
experiments, parameter $i$ will lie in the range
associated with all the simulations that have 
$\Delta C \le \Delta C_{i}$, but the remaining $p-1$ parameters need
not {\it simultaneously} lie in their respective,
similarly computed, single-parameter confidence
ranges, {\it in the same set of experiments}. 
We also repeated the
model 1 simulations with
(1) $N_{H}$ increased by a factor of 20, to
$ 10^{23} \rm \ cm^{-2}$, and (2) exposure time reduced by a factor of 10,
and confirmed similar results.

The results for model 2, which includes a narrow Gaussian emission line at
6.4 keV (intrinsic width, $\sigma_{\rm Fe}$ = 0.1 keV, 
equivalent width, EW = 100 eV), are shown in Figure 1b. The energy resolution
of the \asca SIS at 6.4 keV is $\sim 150$ eV or so, and Figure 1b shows
that $\Delta C_{\rm true}$ and $\Delta C_{i}$ still follow the
$\chi_{6} ^{2}$ and $\chi_{1} ^{2}$ distributions respectively, with
excellent agreement.

Figure  1c shows the results 
(again, as expected) of simulations with the Fe K line equivalent
width increased to 500 eV, the remaining parameters being unchanged
(model 3).  
The relations between the parameter ranges associated with
$\Delta C_{\rm true}$ and $\Delta C_{i}$ for model 3
are shown in Figure 2.
The crosses show the 
$\Delta C_{i}$ plotted against
the best-fitting values of each of the six model parameters, 
including the power-law normalization.
It can be seen that for each parameter
the points are consistent
with a single-valued parabolic function, as expected.
Also plotted are the $\Delta C_{\rm true}$
against the best-fitting parameter values (dots).
It can be seen that the dots lie completely inside the parabolic
curves. If one draws a horizontal line on each plot, corresponding
to a $P\%$ confidence level for $p$ parameters
then all the dots 
below the horizontal line include $P\%$ of all simulations.
Since the dots were computed with $p$ adjustable parameters,
the parameter ranges associated with the dots
are associated with the $P\%$ {\it joint} confidence intervals.
However, since the dots are bounded by the crosses, 
the $P\%$ {\it joint} confidence intervals can be computed from
the {\it single-parameter} $\Delta C$ space by choosing the 
appropriate value of the critical $\Delta C$. The latter is the standard
practice used in actual spectral analysis programs like XSPEC.

In models 4 and 5 the intrinsic width of the emission line was 0.3 keV and
0.01 keV respectively, but the center energies and equivalent widths had the
same values as in model 2. Thus models 4 and 5 test the cases when the 
emission line is broader or narrower than the instrumental energy resolution
respectively, and Figures 1d and 1e show that the 
results agree well with the 
predicted
curves even for the very narrow line.

Figure 1f (model 6, Table 1) shows results for a case in which the localized
spectral feature is not an emission line but an absorption edge,
at 0.8 keV with an optical depth 
at the edge energy of $\tau_{0} = 0.2$. Above the edge the optical
depth is $\tau = 0.2(E/0.8 \rm \ keV)^{-3}$. Such edge features,
due to the ionized Oxygen, have been found in several AGNs (e.g.
see Nandra and Pounds 1992). Spectral fitting of the simulated data
was performed with the edge energy fixed or else the fits would become
unstable (the same can happen with real data).  
The $\Delta C_{\rm true}$  and $\Delta C_{i}$ distributions follow
$\Delta \chi_{4}^{2}$ and $\Delta \chi_{1}^{2}$ respectively, as expected.

\section{
CONFIDENCE INTERVALS FOR EMISSION-LINE EQUIVALENT WIDTH}
\label{lineew}

In practice, when analysing real data with
spectral-fitting programs such as XSPEC,
there is a problem with obtaining the confidence regions
of the equivalent width of emission lines. This is because the equivalent
width depends on the normalization and shape of the underlying
continuum, as well as the line intensity. Thus, the equivalent width
cannot in general be included as one of the model parameters.
A common practice is to obtain the confidence region for the line intensity
and then simply scale this by the best-fitting value of the equivalent
width, as computed from the best-fitting values of the continuum parameters.
We can use the results of our simulations to assess the validity of this
approximation. Figure 3 (solid line) shows the actual distribution in the 
equivalent width of the emission line from the model 3 simulations.
The dotted line shows the distribution of the line intensity, simply
scaled by a single number (in this case,
the input equivalent width of 500 eV),
for a direct comparison. The two distributions are virtually indistinguishable
and we confirmed this for the other emission-line models.
Thus with real data, one can safely derive confidence regions for the
equivalent width of an emission line by simply rescaling the confidence
regions of the intensity of that line with the best-fitting equivalent
width value.

\section{VERY WEAK SOURCES}
\label{weaksource}

We repeated the model 2 simulations, reducing the exposure time to
4 Ks, so that the entire 0.5--10 keV spectrum had less than 4,500 counts.
In this case, when spectral fitting the simulated spectra we fixed the
emission-line energy and intrinsic width so that there were a total of
only four free parameters. Exactly the same procedure would be followed
with real data for weak sources since the fits cannot be easily constrained
so that one is forced to consider the restricted  investigation of 
finding the intensity and/or equivalent width of the line for a given 
center energy and intrinsic width, rather than the more general problem
of trying to constrain all three parameters. The results are shown in
Figure 4a which shows that $\Delta C_{\rm true}$ and the $\Delta C_{i}$ 
are in good agreement with
$\Delta \chi_{4}^{2}$ and $\Delta \chi_{1}^{2}$ respectively, as expected.

We then repeated the model 2 simulations again, reducing the exposure
time further to 400 s (less than 450 counts in the entire spectrum,
contained in 330 bins).
The results are shown in Figure 4b which shows that small
discrepancies are apparent between the predicted and measured
distributions of $\Delta C_{\rm true}$ and $\Delta C$ for the line intensity.
In this case, the line-intensity distribution is peaked at
zero due to many non-detections of the
line since there are so few photons. 
This is illustrated in Figure 5 which shows the distribution
of $\Delta C_{i}$ (crosses) and $\Delta C_{\rm true}$ (dots)
as a function of the
best-fitting line intensity for three exposure times, 40 Ks, 4 Ks and
400 s (panels (a), (b), and (c) respectively). In each case the same
four-parameter model was used. 
It is apparent that for the extreme Poisson limit of 400 s exposure time,
the
deviations between model and data are less than expected since
the line intensity has become a trivial parameter. 
In practice, there is not much cause for concern, since one would
not normally try to even obtain upper limits on the line intensity
from a spectrum with so few counts. 
The discrepancies
between the predicted and observed $\Delta C$ distributions are not so bad for 
large $\Delta C$, which corresponds to cases when the line
is more significant. 
Thus, obtaining 90\% confidence upper limits on the line intensity
is still possible but upper limits for lower confidence levels are 
likely to be somewhat over-estimated.

\section{UPPER LIMITS ON WEAK LINE-EMISSION}
\label{weakline}

Our results so far show that for all practical purposes
the $\chi^{2}_{q}$ region gives the
correct confidence range for the intensity or equivalent
width of an emission line, when the `true' equivalent width is
as small as 100 eV. In cases when the line is not detected,
correct upper limits on the equivalent width may be obtained 
provided there are enough counts in the spectrum. However, we now
consider the situation in which the equivalent width of the line
in the {\it `true'} model is small or zero. We addressed this by 
generating simulated data with model 1 (i.e. {\it no emission line} in the
model) and then fitting the simulated data with model 2
(i.e. {\it including an emission line}),
with the intensity as the only free 
line-parameter (fixing the center energy
and intrinsic width at 6.4 keV and 0.1 keV respectively).
This case, corresponding to no emission line in the `true' model, 
when compared with
the model 2 simulations in \S \ref{results}, then gives information on models
with weak line-emission (i.e. input equivalent width between 0 and 100 eV).
The simulations were 
performed  for exposure times of 40 Ks and 4 Ks. Results for the latter
are shown in Figure 6. It can be seen that there is fair agreement
between $\Delta C_{\rm true}$ and $\Delta \chi_{4}^{2}$ and
between the $\Delta C_{i}$ and  $\Delta \chi_{1}^{2}$.
However, there is a slight departure between
the measured and predicted $\Delta C_{i}$  for the line intensity.
In practice, this will result in a slight overestimate of the upper 
limits on the equivalent width of the line.
For an exposure time of 40 Ks, simulations show an effect of similar
magnitude and for an exposure time of 
400 s, the situation is similar to
the extreme Poisson limit discussed in \S \ref{weaksource}.
Thus we conclude that 
upper limits on the line equivalent width or intensity may be over-estimated
when 
the line is weak or absent in the `true' model.
However, the confidence regions are approximately correct for 
practical purposes.

\section{ CONCLUSIONS} 
\label{conclusions}

We have verified that 
for all practical purposes, the method of generating confidence intervals
for a subset, $q$, of $p$ model parameters, using the $\chi^{2}_{q}$
distribution can still be used even if the model has components
which are narrow compared to the instrumental energy resolution
(such as emission lines and absorption edges). 
We have also investigated the weak-source and 
weak emission-line limits and find the method to work, except for
the extreme Poisson limit when there may be one or less total counts
per energy bin. In this case, equivalent widths of emission lines
may be somewhat over-estimated.

It must be remembered, however, that the $\chi^{2}_{q}$ confidence ranges 
can say nothing of the {\it simultaneous} confidence ranges of
the other $p-q$ parameters. For example, suppose one observes
an active galaxy or X-ray binary and measures the magnitude
of an X-ray reflection continuum component (due to Compton-thick scattering)
and the equivalent width of an iron-K line and quotes, as is
common practice, 90\%
confidence errors for one interesting parameter ($\Delta \chi^{2} = 2.7$).
Now, the relation between the iron-K line equivalent width and
the strength of the reflection continuum can be predicted from a 
theoretical model, so one can determine whether the measured values
are consistent with such a model, within the errors. However,
one-parameter errors (as used , for example, in 
Zdziarski \etal 1996) are inappropriate
since these confidence ranges are not simultaneous.
One must use two-parameter errors in such a case.
 
\vspace{2cm}

Much of this work was done during an extended stay at the
Institute of Space and Astronautical Science (ISAS), Japan, during the
summer of 1993 and two weeks in November 1996. The author
thanks everyone in the X-ray astronomy group at
ISAS for their great hospitality.
The author would also like to thank Peter Serlemitsos, 
Richard Mushotzky,
Andy Fabian, and Andy Ptak
for useful discussions, and Keith Arnaud for generally maintaining
XSPEC and fixing bugs promptly.
The author is very grateful to Dr. W. Cash for correcting some
serious errors in an earlier version of the paper and is also
indebted to an anonymous referee for making some extremely
important points which led to a complete revision of the paper.

\newpage

\section*{ Figure Captions}

\par\noindent 
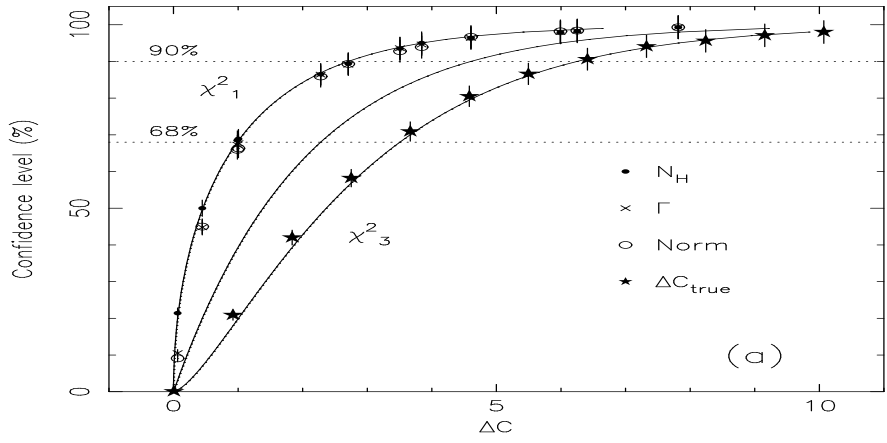
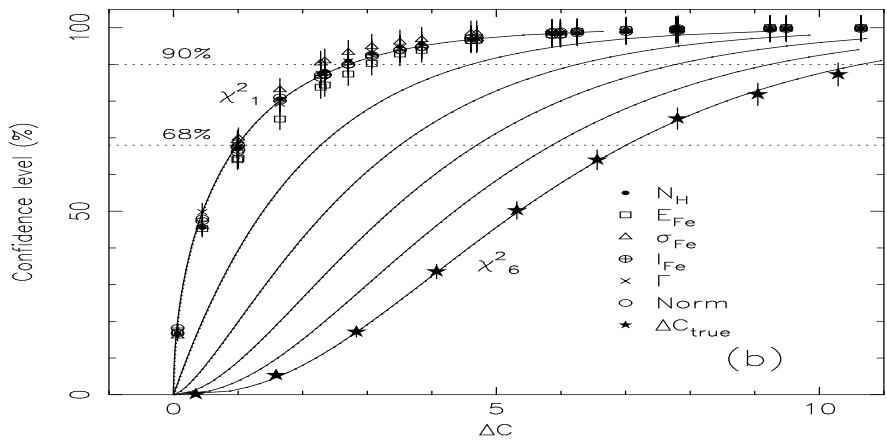
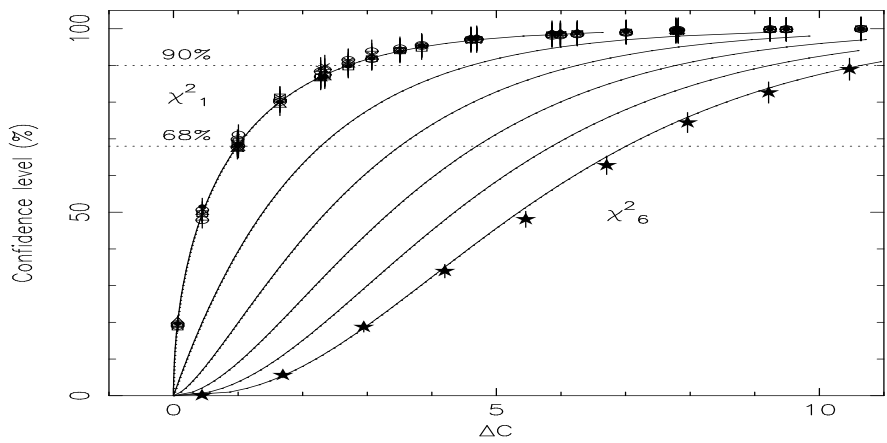
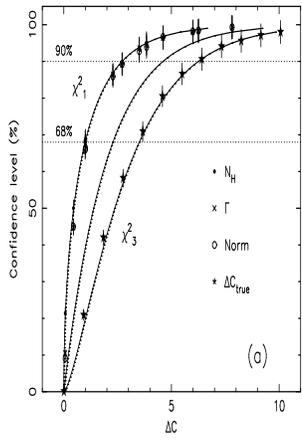
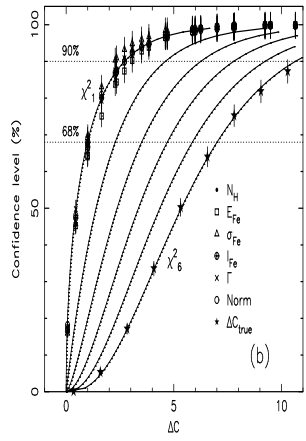
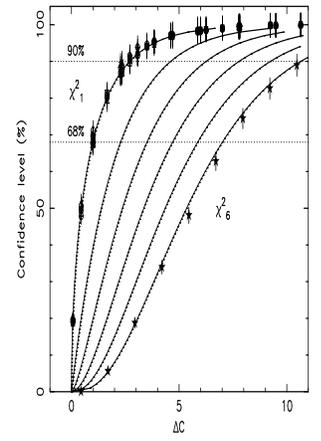
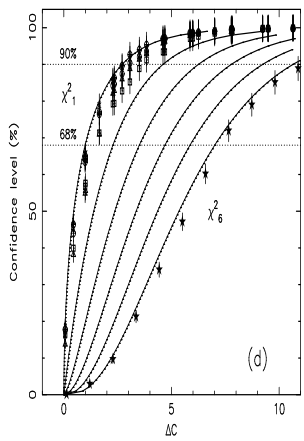
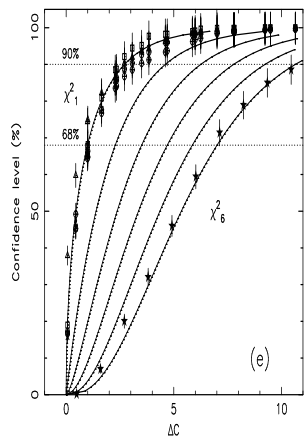
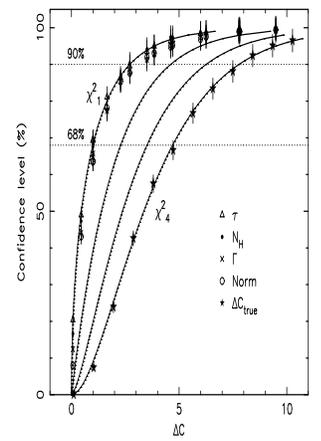
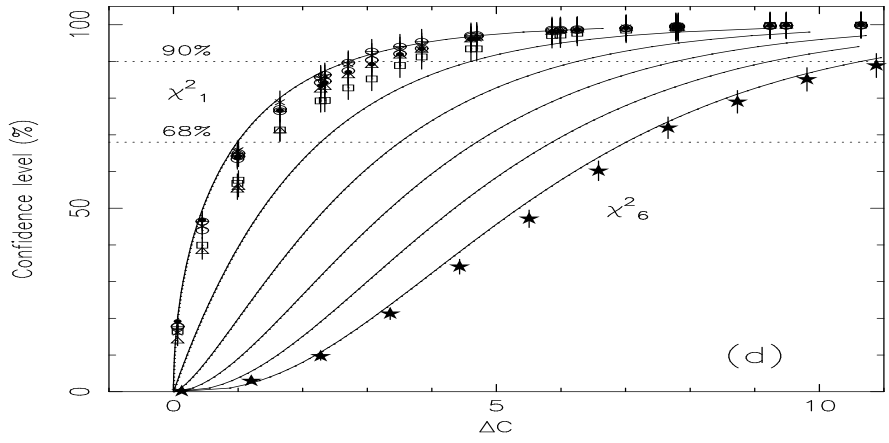
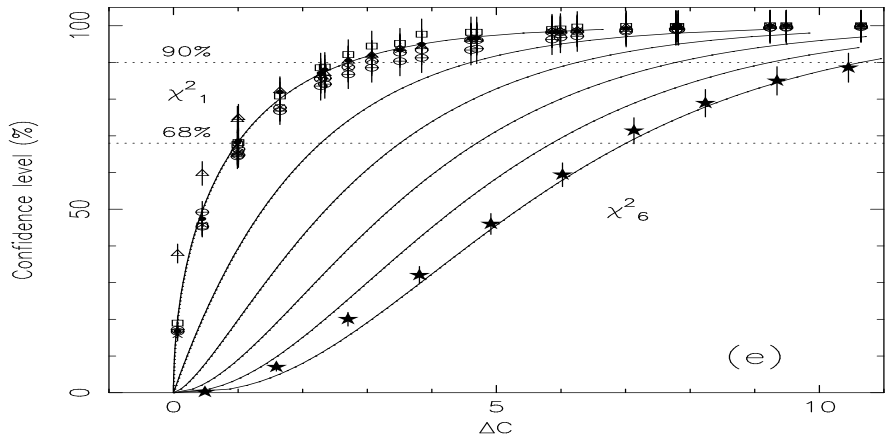
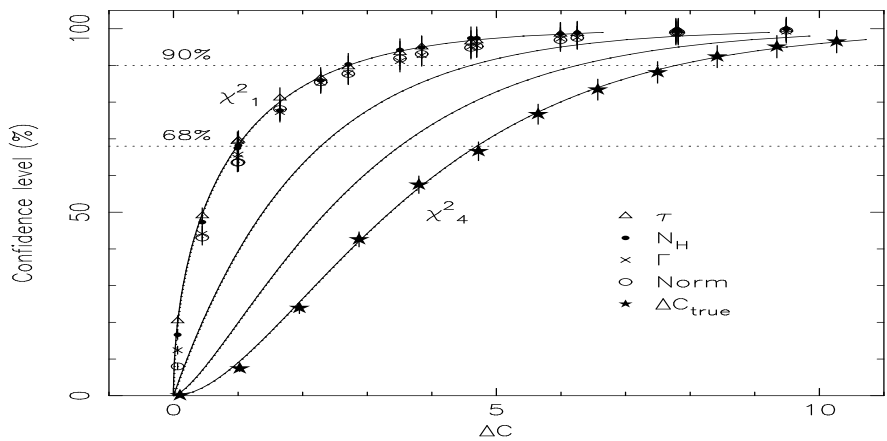
{\bf Figure 1} 

\par\noindent 
Panels (a)--(f) show
results of \asca simulations using models 1--6 respectively
(see Table 1 and text).
Shown are the distributions of $\Delta C_{\rm true} \equiv
C_{\rm true} - C_{\rm min}$ (stars) and $\Delta C_{i} \equiv
C_{i} - C_{\rm min}$ for each
parameter $i$ (the symbols explained in Figures 1a and 1b pertain to
Figures 1c--1e too).
The theoretical $\chi^{2}$ probability distributions (solid curves)
are shown for 1 degree of freedom 
($\chi^{2}_{1}$; uppermost curves) up to 
$p$ degrees of freedom ($\chi^{2}_{p}$; lowest curves),
where $p$ is the total number of free parameters in the model.
In each case the $\Delta C_{\rm true}$
distributions follow $\chi^{2}_{p}$ and the
$\Delta C_{i}$ all follow $\chi^{2}_{1}$, as they should. 

\par\noindent 
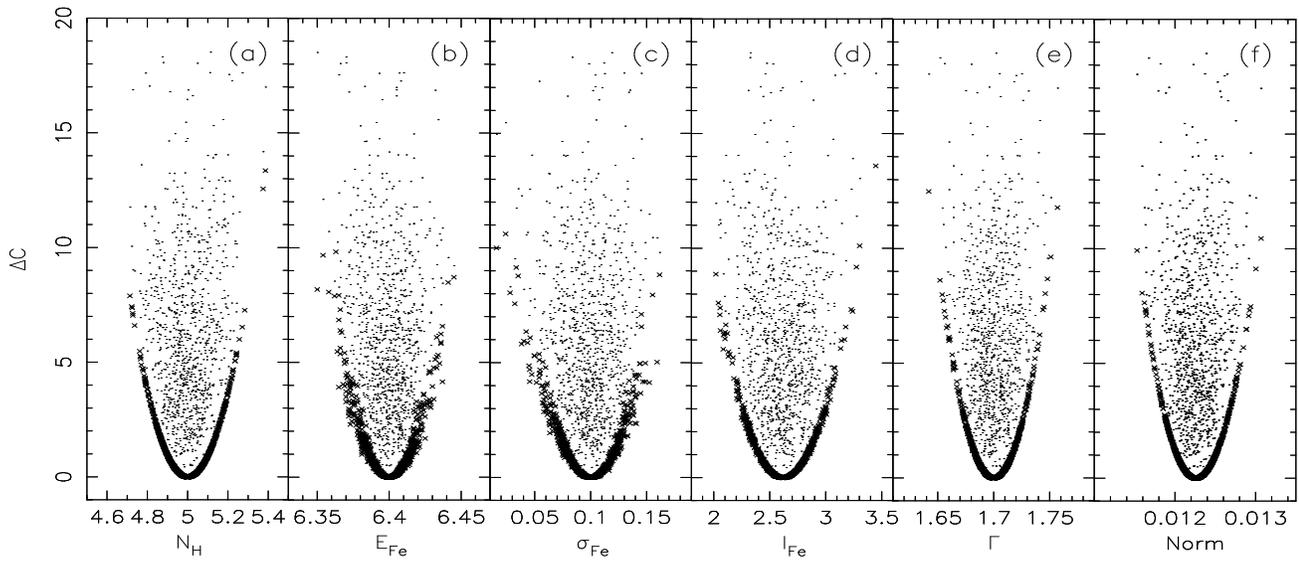
{\bf Figure 2} 

\par\noindent 
Simulation results using model 3 (power law plus a narrow Gaussian
emission line with the parameters shown in Table 1). 
Plots show $\Delta C_{\rm true} \equiv C_{\rm true} - C_{\rm min}$
(dots) and 
$\Delta C_{i} \equiv C_{i} - C_{\rm min}$ (crosses)
versus the best-fitting values of the
model parameters ($N_{H}$ is in units of
$10^{21} \ \rm cm^{-2}$; line center energy, $E_{\rm Fe}$,
and line intrinsic width, $\sigma_{\rm Fe}$ are in units of keV;
line intensity is in units of $\rm 10^{-4} \ photons \ cm^{-2} \ s^{-1}$;
power-law
normalization
is in units  
of 
$\rm photons \ cm^{-2} \ s^{-1} \ keV^{-1}$ at 1 keV;
$\Gamma$ is the power-law photon index).
 
\par\noindent
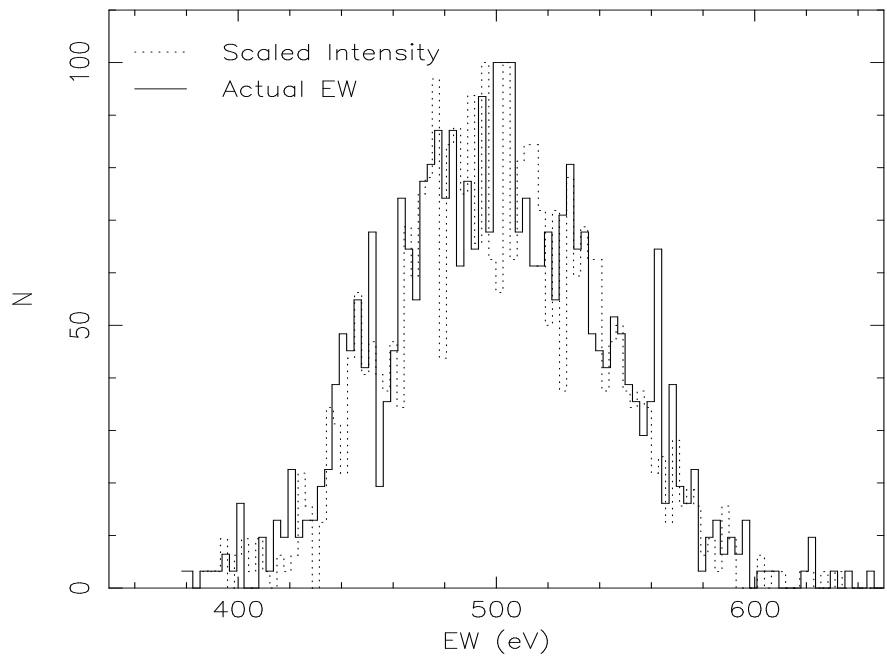
{\bf Figure 3}

\par\noindent
The solid line
shows the distribution
of best-fitting emission-line equivalent width values from the simulations
using model 3 (see \S \ref{lineew}). The dotted line shows the distribution
of best-fitting emission-line intensity values from the same simulations,
scaled by the input equivalent width (500 eV) for direct comparison.
The two distributions are indistinguishable, so for real data, confidence
regions for emission-line equivalent widths can be obtained from
the confidence regions for the intensity, scaled by the best-fitting
equivalent width.
 
\newpage

\par\noindent
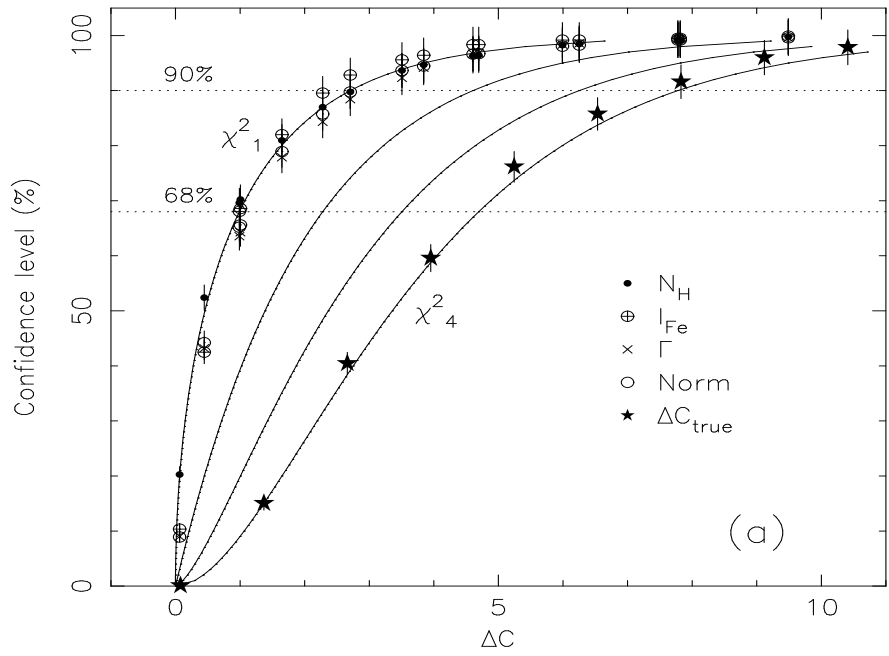
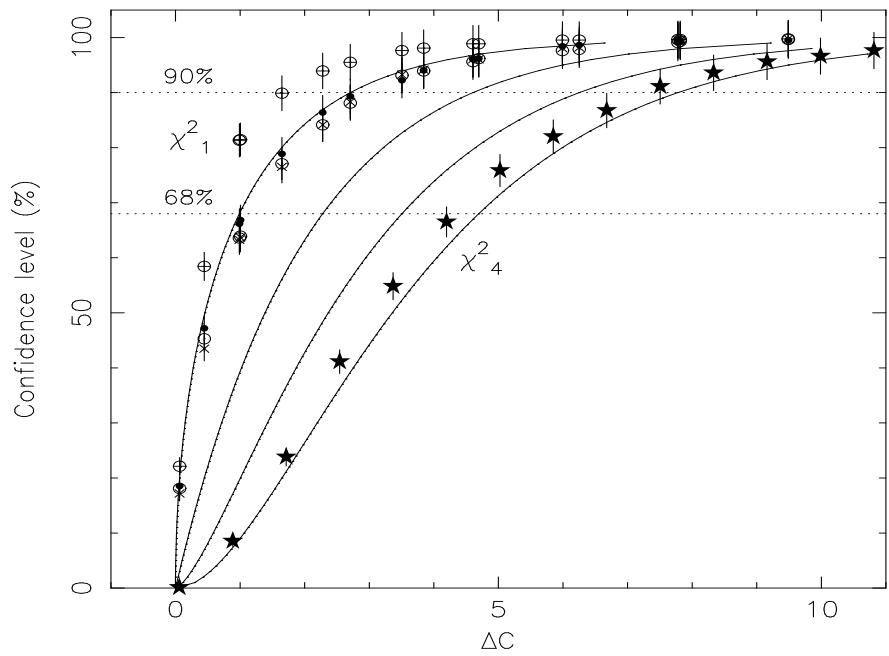
{\bf Figure 4}

\par\noindent
Results of simulations using model 2 (absorbed
power law plus a Gaussian emission line) with reduced
exposure times of (a) 4 Ks, and (b) 400 s.
The center energy and intrinsic width of the Gaussian emission-line
component are held fixed at 6.4 keV and 0.1 keV respectively.

\par\noindent
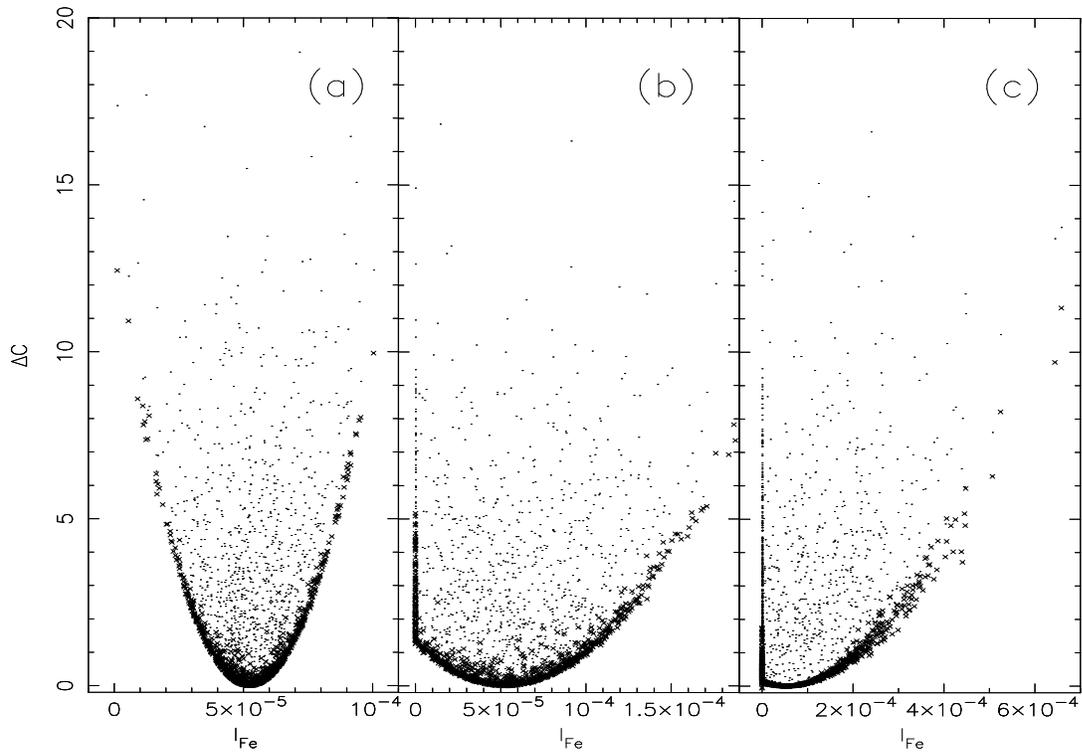
{\bf Figure 5}

\par\noindent
Results of simulations using model 2 (absorbed
power law plus a Gaussian emission line) with 
exposure times of (a) 40 Ks, (b) 4 Ks, and (c) 400 s.
Shown are the distributions of 
$\Delta C_{\rm true} \equiv C_{\rm true} - C_{\rm min}$
(dots) and
$\Delta C_{i} \equiv C_{i} - C_{\rm min}$ (crosses)
versus the best-fitting values of the emission-line intensity, $I_{\rm Fe}$, 
in units of $\rm photons \ cm^{-2} \ s^{-1}$.

\par\noindent
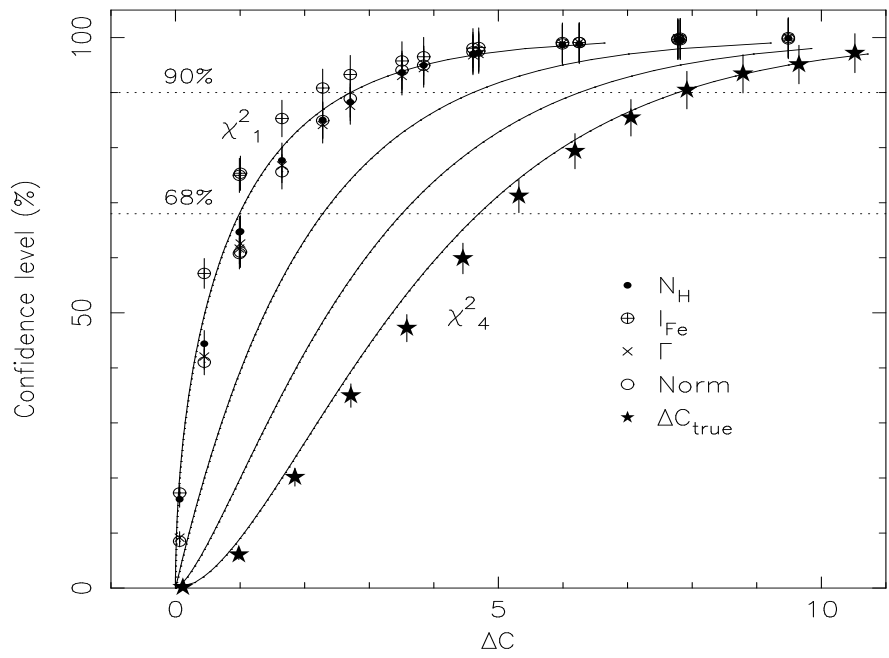
{\bf Figure 6}

\par\noindent
Results of simulations using model 1 (simple absorbed power law only),
in which the simulated spectra are fitted with an additional Gaussian
emission line in the model. This simulates the case of obtaining
confidence regions (upper limits)
on the intensity or equivalent width of an emission
line when the `true' model has no emission line. 
The center energy and intrinsic width of the Gaussian emission-line
component are held fixed at 6.4 keV and 0.1 keV respectively.
See \S \ref{weakline} for details.

\newpage

\begin{table}[h]

\begin{center}
TABLE 1: MODELS USED IN THE SIMULATIONS
\vspace{1cm}

\begin{tabular}{llc}
\hline
& & \\
Model & \hspace{4cm} Description  & Free \\
& & Parameters \\
& & \\
\hline
& & \\
1 & Power law  plus absorption (photon index, $\Gamma =1.7$, & \\
& normalization $=1.226 \times 10^{-2} \rm \ photons \ cm^{-2} \
s^{-1} \ keV^{-1}$ @ 1 keV,  & \\
& column density, $N_{H} = 5 \times 10^{21} \ \rm cm^{-2}$). &  3 \\
& & \\
2 & As model 1, with the addition of a Gaussian line with & \\
& center energy, $E_{\rm Fe}= 6.4$ keV, equivalent width, $EW = 100$ eV & \\
& (intensity, $I_{\rm Fe} = 5.2356 \times 10^{-5} \rm \ photons \ cm^{-2} \
s^{-1}$) & \\
& and intrinsic width, $\sigma_{\rm Fe}= 0.1$ keV
(FWHM $\sim 2.35 \sigma_{\rm Fe}$). & 6 \\
& & \\
3 & As model 2 except that $EW = 500$ eV. & 6 \\
& & \\
4 & As model 2 except that $\sigma_{\rm Fe} = 0.3$ keV. & 6 \\
& & \\
5 & As model 2 except that $\sigma_{\rm Fe} = 0.01$ keV. & 6 \\
& & \\
6 & Power law  plus absorption (photon index, $\Gamma =1.7$, & \\
& normalization $=1.226 \times 10^{-2} \rm \ photons \ cm^{-2} \
s^{-1} \ keV^{-1}$ @ 1 keV,  & \\
& $N_{H} = 5
\times 10^{21} \ \rm cm^{-2}$) plus an absorption edge at 0.8 keV  & \\
&  (fixed) with an optical depth
at threshold, $\tau_{0} = 0.2$. & 4 \\
& & \\
\hline
\end{tabular}
\end{center}
\end{table}


\newpage

\begin{references} 


\par\noindent
\reference{avin76} Avni, Y. 1976, ApJ, 210, 642 
\reference{cash76} Cash, W. 1976, A\&A, 52, 307 
\reference{cash79} Cash, W. 1979, ApJ, 228, 939
\reference{lamp76} Lampton, M., Margon, B., \& Bowyer, S. 1976, ApJ, 208, 177
\reference{nand92} Nandra, K., \& Pounds, K. A. 1992, Nat, 359, 215
\reference{sha89} Shafer, R. A., Haberl, F., \& Arnaud, K. A.  1989, XSPEC: An
X-ray Spectral Fitting Package (ESA TM-09)
\reference{Tana1994} Tanaka, Y., Inoue, H., \& Holt, S. S. 1994, PASJ, 46, L37
\reference{zdzi1996} Zdziarski, A. A., Johnson, W. N., \& Magdziarz, P. 
1996, MNRAS, 283, 193

\end{references}
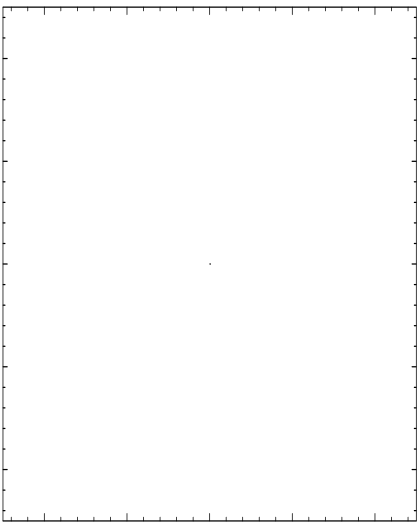
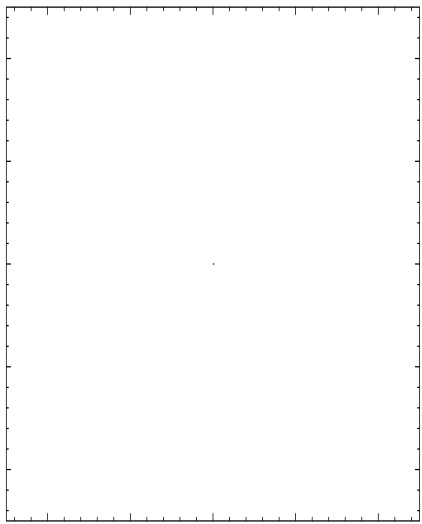
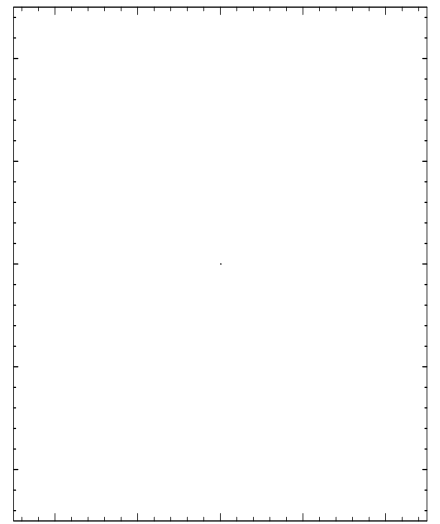
\end{document}